\documentclass{appolb}
\usepackage{graphicx}
\usepackage{amsmath}
\bibliography{references.bib}

\begin{document}

\title{Estimating at Earth the Ultra-High Energy Neutrino Flux from the Accretion Disks in the Galactic Core}
\author{A. Banes\footnote{Presenter of abstract 55 (2), 208.07 at The 241st meeting of the American Astronomical Society, Seattle, WA, 8-12 Jan. 2023, and this project is the topic of the Honors College Senior Thesis.}, T. English and N. Solomey
\address{Physics Division, Wichita State University, KS USA 67260}}

\maketitle
\begin{abstract}
The purpose of this research is to determine at Earth the high-energy neutrino flux coming from the galactic core, Sagittarius A$^*$ (Sgr A$^*$) and from the many other accretion disks within the galactic core. It is estimated there are 10,000 such accretion disk within the cubic parsec of the galactic core alone and many more in the galactic core. There are various neutrino detectors, such as IceCube, which can detect energetic neutrinos. However, the direct galactic core neutrino flux is exceptionally low, so very few neutrinos from the galactic core are measured. We created two models to simulate the galactic core neutrino flux. The first model is a simple linear simulation that predominately relied on the properties of the accretion disks and Sgr A$^*$, which included the quantity, sizes, and distances of the accretion disks. To better estimate the neutrino flux, we replaced the linear accretion disks distribution with a more robust code that randomly distributed the accretion disks and generated bodies of varying sizes. This was then used to determine the ultra-high energy neutrino flux to be $1.80 \times 10^{-17} cm^{-2} s^{-1} sr^{-1} MeV^{-1}$. Since it is extremely difficult to determine neutrino direction from interactions of neutrinos, we envision an application where the energetic galactic core neutrinos are gravitationally focused by the Sun with a “light” collecting power of 10$^{11}$--10$^{12}$. They could interact in a planet's atmosphere where the produced showers containing energetic charged particles can produce Cherenkov rings imageable by an orbiting spacecraft or upward going muons which can be observed in a cosmic ray experiment. Estimating the flux will provide a general approximation of the number of ultra-high energy neutrinos that should reach Earth, with Earth being the overall detector. Moreover, studying the neutrinos will provide more information on the conditions of the galactic region, and allow characterisations of it to be formed.
\end{abstract}
\newpage
  
\section{Introduction}
\begin{figure}[htb]
\centerline{%
\includegraphics[width=12.5cm]{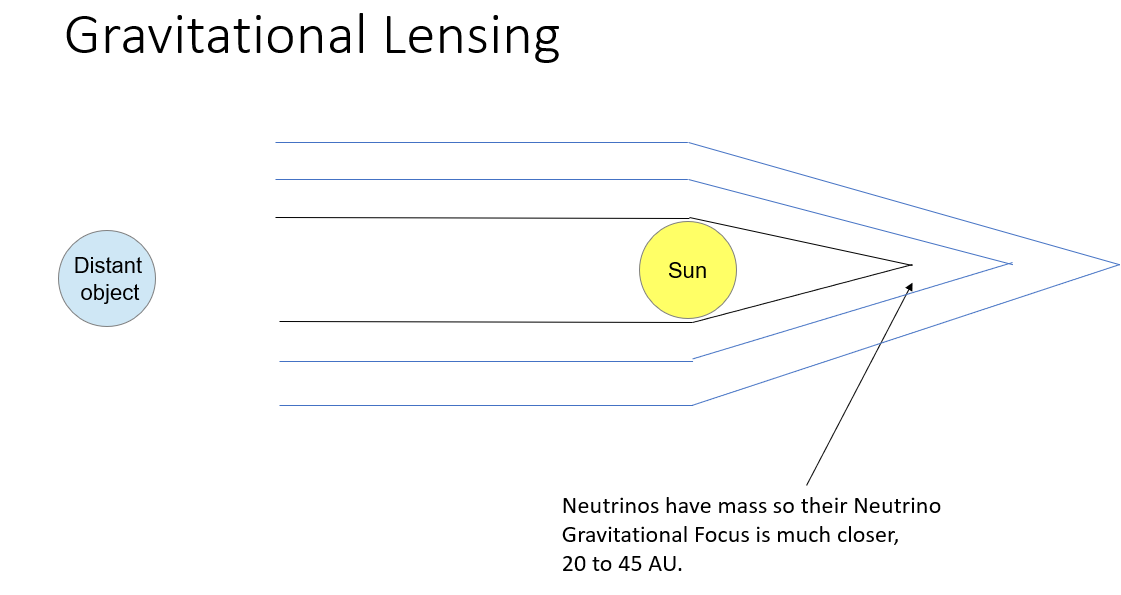}}
\caption{Gravitational Lenzing of a distant object by the mass of the Sun.}
\label{Fig:ngf}
\end{figure}

Located 25,800 light years away, a supermassive black hole (around 4 million times the mass of the Sun) named Sagittarius A* (Sgr A*) lies in the center of the Milky Way galaxy. Sgr A* is one of the many sources for neutrino emission, which are low mass, weakly interacting particles that travel near the speed of light. Due to the high-energy environment, the primary neutrinos emitted are ultra-high energy from 10$^9$-10$^{18}$ eV, i.e. GeV to EeV.\cite{Caballero12} However, the neutrino flux of the galactic core is still unknown. In prior studies Lynn Buchele, with physics professor Dr. Nick Solomey supervising, performed a Senior Physics Project that studied the flux of stellar fusion neutrinos from all the stars in the galactic core and central core of our Galaxy.\cite{Buchele19} If visible at earth to the naked-eye, then the galactic core would be only two times larger than our moon in the night sky. Therefore, it is reasonably small to consider imaging it with a gravitational lens. Although, if the gravitationally lensing of the sun were used, then the neutrinos could be imaged using that method\cite{Einstein36}, which is depicted in figure \ref{Fig:ngf}. Since neutrino have a non-zero mass, this neutrino gravitational lens is expected to be between the orbital distance of Uranus and Neptune\cite{ngf}. Their calculations showed that the measurement of the rate of stellar neutrinos at Neptune and Uranus focused by the Sun will be 800-8000 times greater than the solar fusion neutrino rate of our Sun's direct neutrino flux at Earth because of the large "light" collecting power of a gravitational lens.\footnote{This calculation used the distribution of stars in the galactic core by the model of Kent 1991. It modified each star's neutrino emission by its mass relative to the Sun's mass, and its known fusion neutrino emissions with a further correction for the distance to the star from Earth.} Although, this previous estimate did not encompass the galactic core as a high-energy neutrino source. Other work has been done to simulate high-energy neutrinos on the dark side of an Ice-Giant planet producing energetic showers that could be imaged by an orbiting spacecraft.\cite{English-Paris22,English22}

Accretion disks are circumstellar disks that mainly consist of the material from the object it is orbiting, and are known to excrete a considerable amount of ultra-high energy particles including neutrinos \cite{Spier13}. Over time, as accretion disks accumulate more mass, it can change sizes and increases the number of particle interactions.\cite{Hawkins06} This results in the energy and temperature to rise, which then accelerates the disk and advances the probability of energetic particles to be produced from the high-energy interactions. If the energy of the particle is high enough, then particles can escape the disk; therefore, increasing the neutrino emissions. A famous supernova remnant is Messier 1, also known as the Crab Nebula, which is one of the most studied objects in space. Within the initial attempts of this study, this supernova remnant was treated as a standard candle. All accretion disks in the model were the exact size as the Crab Nebula, and it was used as a reference distance to scale the disks to find our galactic core emission estimate.

This research will attempt to determine the overall flux of ultra-high energy neutrinos produced by the galactic core and the accretion disks within the galactic core. Estimating the flux will provide a general approximation of the number of ultra-high energy neutrinos that should reach Earth, with Earth being the overall detector. Moreover, studying the neutrinos will provide more information on the conditions of the galactic region, and allow characterisations of it to be formed.

\section{Original Attempts}
The first attempt at determining the high-energy accretion disk neutrino flux consisted of using the equation\cite{Surman04}:

\begin{equation}\sigma_e(E_e,Q)= \frac{1}{64\pi}\left(\frac{g_w}{M_w c^2}\right)^4(\hbar c^2)(c^2_V)
\end{equation}

\hspace{-6mm}with g$_W$ as the dimensionless weak coupling constant, M$_W$ is the mass of the W boson, M$_W$c$^2$ = 80.9 GeV, and c$_V$ = 1 as the vector coupling. The primary issue with this attempt were the undefined variables. Q was undefined, and assumed to be the binding energy of the different elements in the accretion disk. However, this would have generated a complicated model because it required knowing the energies and binding energies of each element at the various radii of each accretion disk.

The second attempt involved an updated version of the first attempt, but instead, it was dependent on energy and temperature\cite{Caballero12b}.

\begin{equation}
    \phi(E_{em}) = \frac{g_\nu c}{2\pi^2(\hbar c)^3}\frac{E^2_{em}}{exp(\frac{E_{em}}{T_{em}})+1}
\end{equation}

\hspace{-6mm}This equation was also difficult to utilize because of the constant changing of temperature and energy throughout the disks, as well as the location of the disks within the galactic region. 

In the third attempt, the equation\cite{Bhadra09} shown below was promising in calculating the flux. However, after unit analysis, it was discovered that the variables did not result in units of flux. 

\begin{equation}
 \phi ~= 2c \xi \zeta \eta f_b f_d (1-f_d)n_o \left(\frac{R}{d}\right)^2 P_c
 \end{equation}

\hspace{-6mm} Due to the issues and complexity of the equations, it was determined that creating a model based on the limitations known of the galactic core and of the accretion disks will compute an estimation of the ultra-high energy neutrino flux. This will allow for complete control of all the variables and classify what was more important for the study, such as adjusting the number of accretion disks and their disk sizes.

\begin{figure}[htb]
\centerline{%
\includegraphics[width=12.5cm]{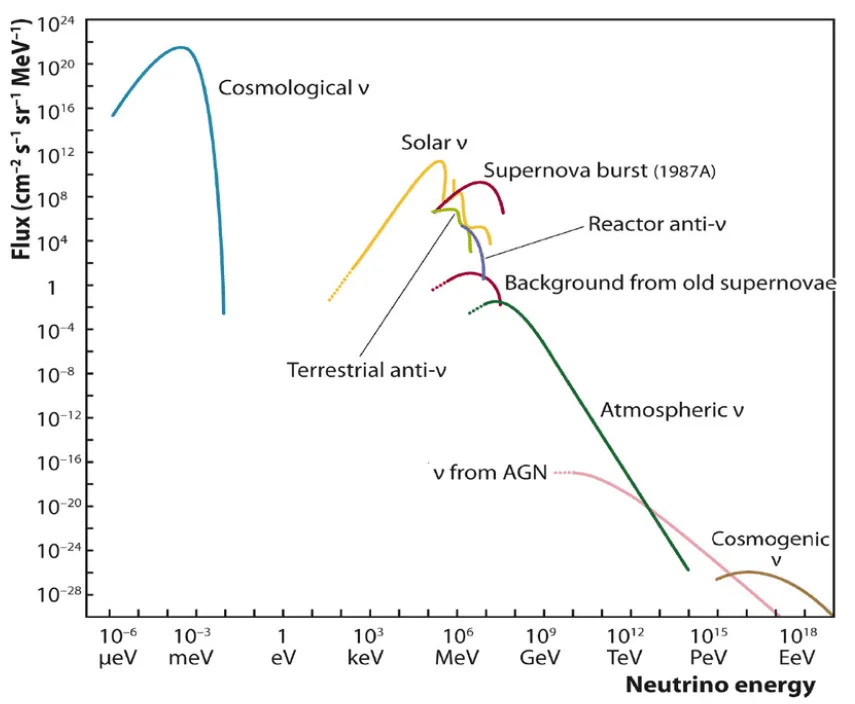}}
\caption{Expected flux of galactic neutrinos from different sources as a function of energy in eV.\cite{Spier13}}
\label{Fig:gn}
\end{figure}

\section{Methods}
Using the program MATLAB, a model of a simple linear distribution simulation that relied on the quantity, sizes, and distances of the accretion disks and Sgr A* was created. To formulate the model, the Crab Nebula was used as a standard candle reference for all accretion disks within the galactic core. The distance from the Earth to the nebula is about 2 kpc, and the Crab Nebula's flux is measured to be $ 8\cdot10^{-15} cm^{-2}s^{-1}$\cite{Bhadra09}, emitted in 4$\pi$ sr or $ 6.37\cdot10^{-16} cm^{-2}s^{-1}sr^{-1}$ Assuming that there are 5,400 accretions disks in the galactic core that are the same size as the Crab Nebula, they were placed in a straight line that was logarithmically spaced over the distance of the galactic region, which is 2-3 kpc.\cite{Weber,Torne21,Zoccali16} Towards the center of Sgr. A*, the disks will be closer together and gradually get further from one other as it nears the edges of the core. 

Accordingly, the distances of each accretion disk from the galactic core to Earth were determined through Pythagorean Theorem. Scaling the distance from each disk to the Crab Nebula, provided an estimate of the flux from each accretion disk and found the total sum by using the equation below.

\begin{equation}
    \phi_{AD} = 2 \sum{\frac{\phi_{CN}}{d_{scaled CN}}} 
\end{equation}

We repeated the same method to determine the neutrino flux coming from Sgr. A* with the assumption of a galactic core flux of $10^{-16} cm^{-2} s^{-1} sr{-1}$ from the minimum expected of an Active Galactic Nuclei (AGN) shown in Figure 2. This is at the distance of 25,800 ly and scaling it allowed for a proper correction in the reduction in signal. Adding the summations together, and doubling them due to symmetry, provided the total flux for this simulation. However, to more accurately estimate the neutrino flux, the linear accretion disk distribution was replaced with a more robust code.

Rewriting the foundation of the linear code in C++ allows for the randomization of the accretion disk placements and sizes. With the previously assumed number of 5,400 accretion disks, each disk had a random generation of 3-dimensional Cartesian coordinates within the galactic core. The clustering of disk closer to Sgr A* and the sparseness as the disks were further away from the center remained the same. Secondly, the the accretion disks sizes were randomized with a lower limit of 0 light days to an upper limit of 10 light days (about $8.3943 \cdot 10^ {-6} kpc$) \cite{Hawkins07}. 

Having gathered the randomized coordinates of each accretion disk and assuming that $x_0 = y_0 = z_0 = 0$ are the Earth's coordinates, the distance from Earth to the disk were able to be calculated using the third-dimensional distance formula below.

\begin{equation}
    d = \sqrt{x^2+y^2+z^2}
\end{equation}

 All the distances to each disk were then scaled to the distance and flux of the Crab Nebula, which then approximated the flux for all the disks in terms of distance. Due to the Crab pulsar (the star in the center of the Crab Nebula) being relatively young, it was assumed that the accretion disk would be around 3 light days. Scaling the randomized accretion disks radius to the assumed radius of the pulsar provided another approximation of the flux in terms of the radius. By taking the average of the flux in terms of distance and radius of all 5,400 accretion disks, the estimated ultra-high energy neutrino flux was determined. 

\section{Results}
All prior attempts had issues that made a successful equation puzzling to solve. Either the equation required a simulation that consisted of accounting for an excess of information or the equation itself did not have the proper units of flux. 

The linear simulation provided a preliminary neutrino flux of the galactic core, and showed that the probability of ultra-high energy neutrinos reaching Earth is scarce. This result does not incorporate any form of special relativity but is simply the flux of the galactic core if no object or phenomenon was between it and the Earth. Although, if the light collecting power of gravitational lensing was included, it will amplify the brightness by at least $10^{11}$--$10^{12}$\cite{Turyshev20}. It is also possible to improve this by two orders of magnitude or more ($\sim10^{13}$) because the limb of the Sun is transparent to neutrinos below 75 GeV. This adds a substantial area for enhancing the "light" collecting power of a gravitational lens for lower energy neutrinos, but not for the far higher ultra high-energy neutrinos. 

By randomizing the accretion disks placements and sizes, the simulation became a more reasonable model of the galactic core. Through all the calculations, it was determined that the ultra-high energy neutrino flux is: $1.8 \times 10^{-13} cm^{-2} s^{-1} sr^{-1}$. Neutrinos from active galactic nuclei (AGN) have a vast amount of energy due to the high energy conditions they are born in. Since the neutrinos from an AGN are expected to be peaked at 10,000 MeV, see figure \ref{Fig:gn} then this is $1.8 \times 10^{-17} cm^{-2} s^{-1} sr^{-1} MeV^{-1}$. Although, the expected flux is right on the edge of the expected range of $10^{-16} - 10^{-28} cm^{-2} s^{-1} sr^{-1} MeV^{-1}$ \cite{Spier13}. Although this was just a first guess toy model it puts us right at the expected edge and if there are ten times less accretion disks then it would improve our match to expectations. There are many reasons from technical, to science, to explain why the real value might be lower than the estimate here:
\begin{enumerate}
    \item The neutrino flux from or the distance to the Crab Nebula source used here as a standard candle is wrong. But our estimate can be scaled by using an improved Crab Nebula number or another standard source.
    \item There are far fewer accretion disks in the galactic core than currently expected.
    \item The accretion disks randomly distributed were assumed to all be identical to the Crab Nebula source and a flat distribution of different size sources each equally possible scaled. More likely, larger accretion disks are exponentially rarer than smaller ones, with these smaller disks producing less and lower energy neutrinos which would need to be entered into the simulation and recalculated.
    \item Neutrino propagation to the Earth is assumed to be unaffected by matter. But if more stars, neutron stars, black holes, or dark matter could interact and stop neutrinos than expected, while dark matter interaction might even do the reverse in increase neutrino flux along its path, none of which has not been taken into account in our estimation. These effects could be an exciting application of this idea to study the space over-which the neutrino propagates from the Galactic Core to Earth.
\end{enumerate}

\section{Conclusion}
Studying the ultra-high neutrinos from the galactic core will not only inform researchers of the probability of detecting them, but it will also allow an expanse of knowledge about the characteristics of Sgr A* and the environment around it. Neutrinos rarely interact with ordinary matter, therefore, if a galactic neutrino is detected, its properties will be similar to how it was within the galactic region. Additionally, due to their low mass, neutrinos always travel near the speed of light, which will provide a relatively recent analysis of the activities occurring in the center of the galaxy.

\section*{Acknowledgments}
This idea stemmed from the interaction of Prof. N. Solomey at a NASA Innovative Advanced Concept Fellow (NIAC) discussion with other NIAC fellows at the 2018 NIAC Symposium; especially with Dr. Slava Turyshev from NASA's JPL Lab who is working on gravitationally focusing light to image an exoplanet.\cite{Turyshev20}

\end{document}